\documentclass[11pt]{article}
\usepackage[margin=1in]{geometry}
\usepackage{natbib}
\usepackage{ragged2e}
\usepackage[labelfont=bf,size=footnotesize]{caption}

\usepackage{xcolor}
\usepackage{hyperref}
\definecolor{urlcolor}{rgb}{0, 0.498, 0.675}
\definecolor{linkcolor}{rgb}{0, 0.498, 0.675}
\definecolor{citecolor}{rgb}{0, 0.498, 0.675}
\hypersetup{breaklinks, colorlinks=true, citecolor=citecolor, linkcolor=linkcolor, urlcolor=urlcolor}

\usepackage{scalerel}
\usepackage{tikz}
\usetikzlibrary{svg.path}
\definecolor{orcidlogocol}{HTML}{A6CE39}
\tikzset{
  orcidlogo/.pic={
    \fill[orcidlogocol] svg{M256,128c0,70.7-57.3,128-128,128C57.3,256,0,198.7,0,128C0,57.3,57.3,0,128,0C198.7,0,256,57.3,256,128z};
    \fill[white] svg{M86.3,186.2H70.9V79.1h15.4v48.4V186.2z}
                 svg{M108.9,79.1h41.6c39.6,0,57,28.3,57,53.6c0,27.5-21.5,53.6-56.8,53.6h-41.8V79.1z M124.3,172.4h24.5c34.9,0,42.9-26.5,42.9-39.7c0-21.5-13.7-39.7-43.7-39.7h-23.7V172.4z}
                 svg{M88.7,56.8c0,5.5-4.5,10.1-10.1,10.1c-5.6,0-10.1-4.6-10.1-10.1c0-5.6,4.5-10.1,10.1-10.1C84.2,46.7,88.7,51.3,88.7,56.8z};
  }
}

\newcommand\orcid[1]{\,\href{https://orcid.org/#1}{\raisebox{2pt}[0pt][0pt]{\scalerel*{
\begin{tikzpicture}[yscale=-1,transform shape]
\pic{orcidlogo};
\end{tikzpicture}
}{+}}}\,}

\newcommand{\nocitep}[1]{}
\newcommand{\nocitealp}[1]{}
\newcommand{\noshow}[1]{}

\begin{document}


{\bf\Large SED-ML Validator: tool for debugging simulation experiments\par}

{\RaggedRight
Bilal Shaikh\orcid{0000-0001-5801-5510}$^{1,*}$, 
Andrew Philip Freiburger\orcid{0000-0002-7288-535X}$^{2,*}$, 
Matthias K\"onig\orcid{0000-0003-1725-179X}$^{3}$, 
Frank T. Bergmann\orcid{0000-0001-5553-4702}$^{4,5}$,\\
David P. Nickerson\orcid{0000-0003-4667-9779}$^{6}$, 
Herbert M. Sauro\orcid{0000-0002-3659-6817}$^{7}$, 
Michael L. Blinov\orcid{0000-0002-9363-9705}$^{8}$, 
Lucian P. Smith\orcid{0000-0001-7002-6386}$^{7}$,\\
Ion I. Moraru\orcid{0000-0002-3746-9676}$^{8}$ and 
Jonathan R. Karr\orcid{0000-0002-2605-5080}$^{1,\dagger}$
}

{\RaggedRight
$^1$Department of Genetics and Genomic Sciences, Icahn School of Medicine at Mount Sinai, New York, NY 10029, USA,
$^2$Department of Civil Engineering, University of Victoria, Victoria, BC V8P 5C2, Canada,
$^3$Department of Theoretical Biology, Humboldt University, 10115 Berlin, \\Germany,
$^4$BioQUANT/COS, Heidelberg University, 69120 Heidelberg, Germany, 
$^5$Department of \\Computing and Mathematical Sciences, California Institute of Technology, Pasadena 91125, CA, \\USA,
$^6$Auckland Bioengineering Institute, University of Auckland, Auckland 1010, New Zealand,\\
$^7$Department ofBioengineering, University of Washington, Seattle, WA 98105, USA and
$^8$Center \\for Cell Analysis \& Modeling, University of Connecticut School of Medicine, Farmington, CT \\06030, USA
}

$^*$These authors contributed equally to this work.\\
$^\dagger$To whom correspondence should be addressed: \href{mailto:karr@mssm.edu}{karr@mssm.edu}.



\section*{Abstract}
\setlength{\parindent}{0pt}%
\setlength{\parskip}{5pt}%
\noindent\textbf{Summary:} More sophisticated models are needed to address problems in bioscience, synthetic biology, and precision medicine. To help facilitate the collaboration needed for such models, the community developed the Simulation Experiment Description Markup Language (SED-ML), a common format for describing simulations. However, the utility of SED-ML has been hampered by limited support for SED-ML among modeling software tools and by different interpretations of SED-ML among the tools that support the format. To help modelers debug their simulations and to push the community to use SED-ML consistently, we developed a tool for validating SED-ML files. We have used the validator to correct the official SED-ML example files. We plan to use the validator to correct the files in the BioModels database so that they can be simulated. We anticipate that the validator will be a valuable tool for developing more predictive simulations and that the validator will help increase the adoption and interoperability of SED-ML.

\textbf{Availability:} The validator is freely available as a webform, HTTP API, command-line program, and Python package at \href{https://run.biosimulations.org/utils/validate}{https://\allowbreak{}run.\allowbreak{}bio\allowbreak{}sim\allowbreak{}u\allowbreak{}la\allowbreak{}tions.\allowbreak{}org/\allowbreak{}utils/\allowbreak{}val\allowbreak{}i\allowbreak{}date} and \href{https://pypi.org/project/biosimulators-utils}{https://\allowbreak{}pypi.\allowbreak{}org/\allowbreak{}pro\allowbreak{}ject/\allowbreak{}bio\allowbreak{}sim\allowbreak{}u\allowbreak{}la\allowbreak{}tors-\allowbreak{}utils}. The validator is also embedded into interfaces to 11 simulation tools. The source code is openly available as described in the Supplementary data.

\textbf{Contact:} \href{mailto:karr@mssm.edu}{karr@mssm.edu}

\setlength{\parindent}{0pt}
\setlength{\parskip}{8pt}

\section{Introduction}
Expanded capabilities to predict biological behavior are needed to help engineers design synthetic biological systems and help physicians precisely diagnose and treat disease \citep{carrera2015build, marucci2020computer}. Achieving more predictive models will likely require deep collaboration among large teams of modelers, experimentalists, and clinicians \citep{szigeti2018blueprint, singla2021community, waltemath2011reproducible}.

To facilitate collaboration, the Computational Modeling in Biology Network (COMBINE; \citealp{hucka2015promoting}) developed the Simulation Experiment Description Markup Language (SED-ML; \citealp{waltemath2011reproducible}), a common format for describing simulations. At its core, SED-ML describes individual simulations of individual models. Initially, SED-ML focused on continuous kinetic models described with CellML (\citealp{cuellar2003overview}) and the Systems Biology Markup Language (SBML; \citealp{keating2020sbml}). Recently, we have expanded SED-ML to a broader range of models including spatial, flux balance, qualitative, and rule-based models; a broader range of simulation algorithms such as flux balance analysis (FBA) and asynchronous logical simulation; and additional model languages such as the BioNetGen Language (\citealp{pmid19399430}), the SBML Flux Balance Constraints (\citealp{olivier2015systems}) and Qualitative Models (\citealp{chaouiya2013sbml}) packages, and Smoldyn \citep{shaikh2021runbiosimulations}. 

On top of this core functionality, SED-ML can describe sets of simulations of variants of models, such as an ensemble of stochastic simulations or a parameter scan of a model. In addition, SED-ML can describe how to create tables and figures of simulation results.

Several tools can create SED-ML files, including web applications such as JWS Online \citep{peters2017jws}, RunBioSimulations, and SED-ML Web Tools \citep{bergmann2017sed}. SED-ML files can be executed with several tools such as COPASI \citep{bergmann2017copasi}, iBioSim \citep{watanabe2018ibiosim}, OpenCOR \citep{garny2015opencor}, Tellurium \citep{choi2018tellurium}, and Virtual Cell \citep{moraru2008virtual}. Furthermore, SED-ML files can be published with repositories such as BioModels \citep{malik2020biomodels}, JWS Online \citep{peters2017jws}, and Physiome \citep{sarwar2019model}. More information about these and other SED-ML tools is available at \href{https://sed-ml.org}{https://\allowbreak{}sed-\allowbreak{}ml\allowbreak{}.org}.

However, the utility of SED-ML has been hampered by limited support for SED-ML among modeling software tools and by different interpretations of SED-ML among these tools. For example, we have found that Tellurium can only execute a few of the simulations in BioModels.

To help modelers debug their simulations and to push the community to use SED-ML consistently, we developed a tool that thoroughly validates SED-ML files. The tool is available as a webform, HTTP API, command-line program, and Python API. Here, we articulate how modelers can use the validator, summarize the validation rules the validator checks, describe how the validator communicates issues about SED-ML files, and highlight how we have already used the validator to correct the official SED-ML examples, identify and fill gaps in the SED-ML specifications, and identify bugs in the implementation of SED-ML by multiple software tools. We also outline how we plan to use the validator to correct the SED-ML files in BioModels. The Supplementary data provides more information about the validator and how we are using it to drive convergence around SED-ML.


\section{Methods}
Because SED-ML files typically describe simulations of external model files encoded in languages such as CellML, NeuroML \citep{cannon2014lems}, and SBML, we designed the validator to validate COMBINE archives \citep{bergmann2014combine}. COMBINE archives are zip archives that contain one or more SED-ML files, other files that the SED-ML files reference, an OMEX manifest file that summarizes the contents of the archive, and optionally OMEX metadata files \citep{neal2020open} that capture metadata about the archive and its contents. COMBINE archives can be created with several tools such as CombineArchiveWeb and RunBioSimulations. More information about these tools is available at \href{https://sed-ml.org}{https://\allowbreak{}sed-\allowbreak{}ml\allowbreak{}.org}.

\begin{figure}[!t]
\centerline{\includegraphics{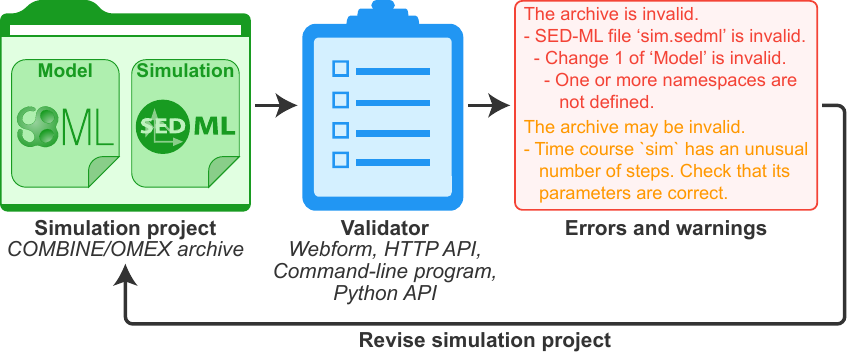}}
\caption{The SED-ML validator helps investigators quickly detect errors and other potential problems in SED-ML and model files organized into COMBINE archives.}\label{fig:overview}
\end{figure}

After creating a COMBINE archive with a SED-ML file, modelers can use the validator via its webform, HTTP API, command-line program, or Python API. Supplementary data S3 and S4 outline how to install and use the validator. To make the validator easy to access, we have also embedded it into standardized interfaces to 11 popular simulation tools (Supplementary data S5.1).

The validator thoroughly checks that COMBINE archives and their contents are consistent with the COMBINE archive, OMEX manifest, OMEX metadata, and SED-ML formats, as well as with the languages of the models in the archive. For example, the validator checks that each reference to a SED-ML element matches the id of an element, that the network of model sources is acyclic, and that each XPath target for each variable of an XML-encoded model matches a single model element. The validator uses LibCellML (\href{https://libcellml.org}{https://\allowbreak{}lib\allowbreak{}cell\allowbreak{}ml.\allowbreak{}org}), LibNeuroML \citep{vella2014libneuroml}, and LibSBML \citep{bornstein2008libsbml} to check that CellML, NeuroML, and SBML models involved in SED-ML files are valid. Supplementary data S2 outlines all of the validation rules that the validator evaluates.

When COMBINE archives are invalid, the validator reports as many errors as can be identified simultaneously, each with contextual information about the element responsible for the error. For example, when the target of a variable of a data generator in a SED-ML file does not match an element of the associated model, the validator provides information about the invalid target and the model that it should match. Supplementary data S4.7 illustrates several example error messages.

The validator also reports warnings about potential mistakes, such as the use of experimental features of SED-ML that few simulation tools support. We implemented warnings for simulations based on common mistakes that we have observed in SED-ML files. We implemented warnings for models using libraries for model languages such as LibSBML.

\section{Real-world examples}
As a first real-world test, we used the validator to identify and fix issues with the official SED-ML examples (Supplementary data S5.3). The validator alerted us to two common problems with each example, as well as less common issues with several files. The validator also identified files that use a combination of SED-ML elements that the SED-ML specifications do not officially support. This finding prompted us to add a warning for this combination of elements and clarify the description of these examples on the SED-ML website. To ensure these examples remain valid, we also set up an automated action that uses our validator to check these files each time they are changed.

Encouraged by this success, we plan to use our validator to identify and fix issues with the SED-ML files in BioModels (Supplementary data S5.4). We anticipate these corrections will enable these files to be simulated with multiple tools, which will increase the utility of the models in BioModels.

To avoid similar errors in the future, we have also submitted several proposals to clarify the specifications of SED-ML (Supplementary data S5.5) and filed numerous bug reports for several software tools that support SED-ML (Supplementary data S5.6). In addition, we aim to help the BioModels Team incorporate our validator into their curation workflow to ensure that BioModels publishes valid SED-ML files going forward.

\section{Discussion}
We believe that our validator will be a key resource for debugging simulation experiments and that this work will push the community to use SED-ML consistently. Taken together, we believe these advancements will increase the community's ability to collaborate on simulation experiments, which we anticipate will foster more sophisticated models.

Once the next version of SED-ML (L1V4) is approved, we plan to expand the validator to SED-ML's new features for additional types of observables and plots, data reductions, and model calibration. We also aim to expand the capabilities of the validator to validate additional types of files that could be included in COMBINE archives, such as PETab and Vega files, two emerging formats for model calibration and data visualization.


\section*{Funding}
This work was supported by the National Institutes of Health [grant number P41EB023912], BMBF LiSyM [grant number 031L0054], and DFG QuaLiPerF [grant number 436883643].

\textbf{Conflict of Interest:} none declared.


\end{document}